\documentclass[ste,twoside]{stefano}
\usepackage{amsmath}
\usepackage{amssymb}
\usepackage{graphics}
\usepackage{rotating}
\usepackage{cite}
\usepackage{color}

\def\makeheadbox{{%
\hbox to0pt{\vbox{\baselineskip=10dd\hrule\hbox
to\hsize{\vrule\kern3pt\vbox{\kern3pt \hbox{  {\sc
quant-ph/0408028} } \hbox{ {\sc Mod. Phys. Lett. A {\bf 19},
2717-2725 (2004) } \hspace*{6.4cm}
{\color{blue}{$\boldsymbol{\Sigma \delta \Lambda}$}} }
\kern3pt}\hfil\kern3pt\vrule}\hrule}%
\hss}}}

\def\0{\mbox{\tiny $0$}}
\def\1{\mbox{\tiny $1$}}
\def\2{\mbox{\tiny $2$}}
\def\3{\mbox{\tiny $3$}}
\def\4{\mbox{\tiny $4$}}
\def\5{\mbox{\tiny $5$}}
\def\6{\mbox{\tiny $6$}}
\def\7{\mbox{\tiny $7$}}
\def\8{\mbox{\tiny $8$}}
\def\9{\mbox{\tiny $9$}}
\def\R{\mbox{\tiny $R$}}
\def\T{\mbox{\tiny $T$}}
\def\A{\mbox{\tiny $A$}}
\def\B{\mbox{\tiny $B$}}
\def\I{\mbox{\tiny $I$}}
\def\II{\mbox{\tiny $II$}}
\def\III{\mbox{\tiny $III$}}
\begin{document}
%
%%%%%%%%%%%%%%%%%%%%%%%%%%%%%%%% PAPER %%%%%%%%%%%%%%%%%%%%%%%%%%%%%%%%%%%%%

\title{\Large  ABOVE BARRIER POTENTIAL DIFFUSION}
%\subtitle{}

\author{
Alex E. Bernardini\inst{1}
Stefano De Leo\inst{2}
%\thanks{Partially supported by the FAPESP grant 99/09008--5.}
%\and
%Gisele C. Ducati\inst{1,2}
%\thanks{Supported by a CAPES PhD fellowship.}
%\and
%Celso C. Nishi\inst{2}
\and Pietro P. Rotelli\inst{3} }

\institute{
Department of Cosmic Rays and Chronology, State
University of Campinas\\
PO Box 6165, SP 13083-970, Campinas, Brazil\\
{\em alexeb@ifi.unicamp.br} \and
Department of Applied Mathematics, State University of Campinas\\
PO Box 6065, SP 13083-970, Campinas, Brazil\\
{\em deleo@ime.unicamp.br}
%{\em ducati@ime.unicamp.br}
%\and
%Department of Mathematics, University of Parana\\
%PO Box 19081, PR 81531-970, Curitiba, Brazil\\
%{\em ducati@mat.ufpr.br}
%\and Department of Cosmic Rays and Chronology, State University of
%Campinas\\
%PO Box 6165, SP 13083-970, Campinas, Brazil\\
%{\em ccnishi@ifi.unicamp.br}
\and
Department of Physics, INFN, University of Lecce\\
PO Box 193, 73100, Lecce, Italy\\
{\em rotelli@le.infn.it}
}

%%%%%%%%%%%%%%%%%%%%%%%%%%%%%%%%%%%%%%%%%%%%%%%%%%%%%%%%%%%%%%%%%%%%%%%%%%%
%%%%%%%%%%%% DATE ABSTRACT PACS % %%%%%%%%%%%%%%%%%%%%%%%%%%%%%%%%%%%%%%%%%

\date{{\em June, 2004}}
%- Revised version:  {\em April, 2004} }
% Warning: Where is the date?

\abstract{The stationary phase method  is applied to diffusion by
a potential barrier for an incoming wave packet with energies
greater then the barrier height. It is observed that a direct
application leads to paradoxical results. The correct solution,
confirmed by numerical calculations is the creation of multiple
peaks as a consequence of multiple reflections. Lessons concerning
the use of the stationary phase method are drawn.}

%%%%%%%%%%%%%%%%%%%%%%%%%%%%%%%%%%%%%%%%%%%%%%%%%%%%%%%%%%%%%%%%%%%%%%%
%%%%%%%%%%%%%%%%%%%%%%%%%%%%%%%%%%%%%%%%%%%%%%%%%%%%%%%%%%%%%%%%%%%%%%%

%%%%%%%%%%%%%%%%%%%%%%%%%%%%%%%%%%%%%%%%%%%%%%%%%%%%%%%%%%%%%%%%%%%%%%%
%%%%%%%%%%%%%%%%%%%%%%%%%%%%%%%%%%%%%%%%%%%%%%%%%%%%%%%%%%%%%%%%%%%%%%%

%\PACS{ {??.??.?} \and  {??.??.?}{}}

% Warning: No PACS code given

%02.10.Hh Rings and algebras
%02.10.Ud Linear algebra
%02.10.Yn Matrix theory

%02.30.Hq Ordinary differential equations
%02.30.Jr Partial differential equations
%02.30.Tb Operator theory

%03.65.-w Quantum mechanics
%03.65.Ca Formalism
%03.65.Ta Foundations of quantum mechanics; measurement theory
%03.65.Xp Tunnelling, traversal time,quantum Zeno dynamics

%12.15.F Quarks and lepton masses and mixing
%14.60.Pq Neutrino mass and mixing

%\offprints{~Stefano De Leo.}

\titlerunning{\sc above barrier potential diffusion}

\maketitle

%%%%%%%%%%%%%%%%%%%%%%%%%%%%%%%%%%%%%%%%%%%%%%%%%%%%%%%%%%%%%%%%%%%%%%%
%%%%%%%%%%%%%%%%%%%%%%%%%%%%%%  SECTION   %%%%%%%%%%%%%%%%%%%%%%%%%%%%%
%%%%%%%%%%%%%%%%%%%%%%%%%%%%%%%%%%%%%%%%%%%%%%%%%%%%%%%%%%%%%%%%%%%%%%

%\section*{I. INTRODUCTION}

The stationary phase method (SPM), first introduced to physics by
Stokes and Kelvin\cite{K887}, provides an approximate way to
calculate the maximum of an integral.  It has in time become a
standard tool in the armory, not only of physicists but
biologists, economists etc. \cite{PBE}. Below we shall briefly
sketch the method. One of its main attractions is the apparent
insignificance of details
 of the integrand with the exception of its phase. Already
 within its description, a series of limitations and assumptions
 are made. While these are known to the experts they are often
 assumed implicitly and tested indirectly a posteriori by the
 success or otherwise of the results obtained.

Recently much interest in the physics community has been stirred
by the results of this method applied to tunnelling times in a
potential barrier \cite{TT}. This has resulted in  predictions of
superluminal velocities, or more precisely to tunnelling times
which, in the so-called opaque limit, are {\em independent} of the
barrier length. Now, while not addressing this question directly
in this paper, we investigate what we consider a simpler but
related problem: The (non-relativistic) diffusion of an incoming
single wave packet with energy spectrum totally above the barrier
height. We first show that a direct application of the SPM
analogous to the tunneling case (energy spectrum below the barrier
height) also leads to surprising, not to say, paradoxical results.
We have then performed numerical calculations which clearly
display secondary reflected and transmitted peaks. This stimulates
the assumption of multiple reflections which when combined with
the SPM yields excellent agreement with our numerical
calculations. The primary lesson that we draw is that the SPM
without additional knowledge such as the number of wave packets
existing is ambiguous and whence meaningless. For diffusion
problems the conservation of probabilities can in principle be
used to eliminate this ambiguity.

Consider a complex integral over an unspecified range of the form
\begin{equation}
\label{eq1} {\mathcal I} \,= \, \int F(k) \, \mbox{d}k = \int
|F(k)| \exp[i\,\theta(k)]\, \mbox{d}k \, \, ,
\end{equation}
for which $|F(k)|$ has a single maximum within the range of
integration at $k=k_{\0}$. If $\theta(k)$ varies sufficiently
smoothly within the interval where $|F(k)|$ is appreciable, we can
expand $\theta(k)$ about the point $k=k_{\0}$ in a Taylor series
\[
\theta(k)=\theta_{\0}+(k-k_{\0})\, \theta'_{\0} +
\mbox{O}[(k-k_{\0})^{\2}]\, \, ,
\]
where
\[ \theta_{\0} \equiv \theta(k_{\0})\hspace*{.5cm}
\mbox{and} \hspace*{.5cm} \theta'{\0}\equiv \left. \frac{\mbox{d}
\theta(k)}{\mbox{d}k}\right|_{\mbox{\footnotesize
$k\,$=$\,k_{\0}$}}\, \, .
\]
If the modulus of $F(k)$ is sufficiently sharply peaked we can
neglect the second and higher order terms in the above series.
This allow us to approximate the integral in Eq.(\ref{eq1}) by
\begin{equation}
{\mathcal I} \,\approx \,\exp[i\,\theta_{\0}] \int |F(k)|
\exp[i\,(k-k_{\0}) \,\theta'_{\0}]\, \mbox{d}(k-k_{\0})\, \, ,
\end{equation}
However, if $\theta'_{\0}$ is large the function of $k$ which is
to be integrated oscillates rapidly and, consequently, this
integral will be practically null. A significant contribution
occurs only when, for appropriate values of any parameters within
$\theta(k)$,
\begin{equation}
\label{tprime} \theta'{\0} = 0\, \, .
\end{equation}

In this study we consider a modulated plane wave and are
interested in the configuration space wave function in one
dimension $x$,
\begin{equation}
\psi(x,t)=\int |F(k)| \exp[i\lambda(k)] \exp[i(kx-Et)] \,
\mbox{d}k
\end{equation}
with $E=k^{\2}/2m$ for non-relativistic quantum mechanics. The
$F(k)$ may be a gaussian or similar modulation function. The total
phase is
\begin{equation}
\theta(k;x,t)= kx - \frac{\, \, k^{\2}}{2m}\,t+\lambda (k)\, \, ,
\end{equation}
and the condition $\theta'_{\0}=0$ then yields the $x$-$t$
dependence of the maximum or peak of $|\psi(x,t)|$. For example,
when $\lambda(k)=0$, we obtain the group velocity result for a
free wave packet
\begin{equation}
x=\frac{\,k_{\0}}{m}\, t\, \, .
\end{equation}
The existence of a $\lambda(k)$ produces a time or space shift
\[
x=\frac{\,k_{\0}}{m}\, t - \lambda'_{\0} = \frac{\,k_{\0}}{m}\,
\left( t - \frac{m \lambda'_{\0}}{\,k_{\0}}\right)
=\frac{\,k_{\0}}{m}\, \left( t - \Delta t \right)\, \, .
\]
%or
%\[\tilde{x} = x +\lambda'_{\0}=\frac{\,k_{\0}}{m}\, t\, \, .\]
It is exactly this type of analysis which leads to a delay time in
the reflection of an incoming wave packet impacting upon a {\em
step potential} when the momentum or energy spectrum is totally
contained  below the step height\cite{CT}. A similar analysis has
been used for tunnelling times\cite{TTbis}.

The standard procedure in these one dimensional potential problems
is to find the stationary (but not normalizable) plane wave
solutions with the appropriate continuity conditions (see below)
and then pass to a normalized wave packet by means of a modulating
function. While the plane waves exist at all times in an infinite
range of $x$, the wave packet is predicted by the SPM to exist for
the incoming wave for say $t<0$ (with appropriate chosen time
origin) while the reflected wave and other waves exist only for
$t>0$. Around $t=0$ we will have interference effects, due to the
simultaneous presence of both incoming and reflected wave, and for
the below barrier case we also have (over this transitory period)
a wave function within the classically forbidden barrier region.

 In the following
figure, we show the potential barrier

\begin{picture}(180,90) \thinlines
\put(50,10){\vector(0,1){68}} \put(44,80){$V(x)$}
\put(2,10){\vector(1,0){148}} \put(154,8){$x$}
\put(7,25){\mbox{\small \sc Region I}} \put(54.5,25){\mbox{\small
\sc Region II}} \put(105,25){\mbox{\small \sc Region III}}
\put(49,0){$0$} \put(99,0){$l$} \thicklines
\put(50,10){\line(0,1){35}} \put(50,45){\line(1,0){50}}
\put(100,45){\line(0,-1){35}}
\put(2,60){.....................................................}
\put(38,42){$V_{\0}$}  \put(-12,60){$E_{\0}$}  \put(230,42){$ V(x)
= \left\{
\begin{array}{lcl}
0 &  & ~~~~~\,\,x < 0 \, ,\\
V_{\0} & ~~~~~ & \,0 <  x <  l \, ,\\
0 &  & \,\,l < x \, ,
\end{array}
\right.$}
\end{picture}

\noindent divided into three regions I ($x < 0$), II ($0<x<l$) and
III ($x>l$). The dotted line indicates the mean energy of the
incoming wave $\Psi_{inc}$,
\begin{equation}
\label{inc} \Psi_{inc}(x,t) =
\int_{_{_{\hspace*{-0.2cm}\sqrt{2mVo}}}}^{^{\, \infty}}
\hspace*{-0.6cm} g(k) \, \exp[i(kx-Et)] \, \mbox{d} k\, \, ,
\end{equation}
with $g(k)$ a truncated gaussian or similar, peaked at $k_{\0}$
($E_{\0}=k^{\2}_{\0}/2m$). Truncation is needed, at least for
small $k$ as indicated in Eq.(\ref{inc}), since we wish to avoid
any tunnelling phenomena. The $x$-dependence of the plane wave
solutions in the three regions are given by
\begin{equation}
\begin{array}{lclclcl}
 \mbox{\small \sc Region I:} &~~~~ & \hspace*{.65cm} x < 0 \, ,& ~~~ &
 \hspace*{.85cm}\exp[ikx] + R(k) \,
 \exp[-ikx]
 & &
 ~~~[\, k=\sqrt{2\,mE} \, \,]\, , \\
\mbox{\small \sc Region II:} & &0 < x < \,l \, ,&  & A(k) \,
\exp[iqx] + B(k) \,
 \exp[-iqx]
 & &
~~~[\, q=\sqrt{2\,m(E- V_{\0})} \, \,] \, ,\\
\mbox{\small \sc Region III:} &   & \,l < x \, ,&  & T(k) \,
\exp[ikx] \, .  &  &
\end{array}
\end{equation}
$R(k)$ and $T(k)$ are the reflected and transmitted amplitudes
respectively. The coefficients $A(k)$ and $B(k)$ are the right and
left going amplitudes in region II. All amplitudes are to be
modulated by the function $g(k)$ eventually. Continuity of
$\Psi(x,t)$ and its derivative at $x=0$ and $x=l$ determines the
coefficients $A$, $B$, $R$ and $T$,
\begin{equation}
\label{cexp}
\begin{array}{lclclcl}
A(k) &=& k (k + q)\, \exp [i\lambda(k) - iql\,]\,/\,\mathcal{D}(k)
\, ,\, \, & &  B(k) &=& k(q -
k)\, \exp [i\lambda(k) + iql\,]\, /\,\mathcal{D}(k)\, , \\
R(k)      &=& (k^{\2} - q^{\2})\,\sin [ql\,] \, \exp
[i\lambda(k)-i\, \mbox{$\frac{\pi}{2}$}]\,/\, \mathcal{D}(k)\, ,
\, \, & & T(k) &=& 2 k q \, \exp[i\lambda(k) - ikl\,]\,/\,
\mathcal{D}(k)\, ,
\end{array}
\end{equation}
where
\[
\mathcal{D}(k)=\left\{4 k^{\2} q^{\2} + \left(k^{\2} - q^{\2}
\right)^{\2}\sin^{\2}[q l\,]\right\}^{\frac{1}{2}}\, \, \,
\mbox{and}\, \, \,\, \, \lambda(k) = \arctan \left\{ (k^{\2} +
q^{\2} )\, \tan[ql\,]\, /\, 2kq \right\}\, \,.
\]
To apply the SPM in what we would call the naive way, we must
multiply each of the above amplitudes by the appropriate plane
wave phases. For example, in the simplest case when  $g(k)$ is a
{\em real function} we obtain
\begin{eqnarray}
\label{theta} \theta_{inc}(k)& =& kx-Et\,,\\
 \theta_{\R}(k) & = &
\lambda(k) - \, \mbox{$\frac{\pi}{2}$} - kx - Et\,,
\nonumber \\
\theta_{\A}(k) & = & \lambda(k) + q (x-l) - Et\,\,,
\nonumber \\
\theta_{\B}(k) & = & \lambda(k) +q (l-x) - Et\,\,,
\nonumber \\
\theta_{\T}(k) & = & \lambda(k) + k (x -l) - Et\, \, .
\end{eqnarray}
 The presence of the phase term $\lambda(k)$  implies a delay
time in the reflected wave analogous to what happens for the step
potential when $E<V_{\0}$. Since the phase of the incoming wave
contains only the plane wave factors, i.e., it is devoid of a
$\lambda(k)$, the incoming peak reaches the barrier at $x=0$ at
time $t=0$ (neglecting interference effects). For the reflected
peak with the above expression for $\theta_{\R}(k)$, we find the
position of the peak of the reflected wave to be at
\begin{equation}
x=\lambda'(k_{\0}) - (k_{\0}/m)\,t\, \, ,
\end{equation}
 with
\begin{equation}
\lambda'(k_{\0})=\left[\, \frac{2}{q} \, \frac{\left(k^{\2} +
q^{\2} \right)k^{\2} q l - \left(k^{\2} - q^{\2}
\right)^{\2}\sin[q l\,] \cos[q l\,]}{4 k^{\2}  q^{\2}
+\left(k^{\2} - q^{\2} \right)^{\2}\sin^{\2}[q
l\,]}\,\right]_{\mbox{\footnotesize $k\,$=$\,k_{\0}$}}\,\,.
\end{equation}
Note that only $x<0$ is physical in this result since the
reflected wave, by definition, lies in region I.

The above expression for the position of the reflected peak
simplifies around the "resonance" values for $k_{\0}$ ($q_{\0}$)
where
\[ \sin [q_{\0} l\,]=0\, \,, \hspace*{1cm}
\mbox{i.e.}\hspace*{.5cm} q_{\0} l = n \, \pi\, \, ,
\]
with $n$ a non-negative integer. Assuming therefore, for
simplicity,
 a sharp spectrum for $g(k)$ peaked at one of these resonance
values
\begin{equation}
\lambda'_{res}(k_{\0})\,\approx\,
\frac{(k_{\0}^{\2}+q_{\0}^{\2})\,l}{2\,q_{\0}^{^{\2}}}\,>\,0 \, \,
.
\end{equation}
This predicts a delay time for the reflected wave given by
\begin{equation}
\Delta t^{res}_{\R}=
\frac{m}{k_{\0}}\,\lambda'_{res}(k_{\0})\approx
\frac{(k_{\0}^{\2}+q_{\0}^{\2})\,m\,l}{2\,k_{\0}q_{\0}^{^{\2}}}\,\,.
\end{equation}
Now consider the corresponding "delay times" for the $A$, $B$ and
$T$ waves. In particular,
\begin{equation}
\Delta t^{res}_{\A}= \frac{m}{k_{\0}}\,\left[ \,
\lambda'_{res}(k_{\0}) - q'(k_{\0})l\,\right] =\Delta t^{res}_{\R}
-\frac{m}{q_{\0}}\, l \approx
\frac{(k_{\0}-q_{\0})^{^{\2}}m\,l}{2\,k_{\0}q_{\0}^{^{\2}}}\, \, .
\end{equation}
This is the delay time at $x=0$. We can calculate a delay time
even for the $B$-wave at $x=0$ since it exists in region II.
However, this is later than the time of the $B$-wave at $x=l$
since it is left-moving i.e. as a consequence of the $q(l-x)$
factor in Eq.(\ref{theta}). Suffice it to say that it arrives at
$x=0$ at a later time than the departure times of either the $R$
or $A$ wave peaks. For completeness the transmitted $T$ wave
packet has its peak at the start of region III, $x=l$, at the time
\[
t^{res}_{\T}=t^{res}_{\R}\, \, .
\]

Now already one may be somewhat surprised to note that the
appearance of the transmitted wave coincides with that of the
reflected wave. However, the above results become paradoxical as
soon as one realizes that for the time interval from $t=0$ to
$t=\Delta t^{res}_{\A}<\Delta t^{res}_{\R}$ this solution is
devoid of {\em any} maximum ($R$, $A$, $B$ or $T$). During this
time, at least, we are clearly in contradiction with probability
conservation since the incoming wave peak has disappeared at time
$t=0$. By choosing the wave packet dimensions small enough we can
say that there is an interval of time in which the naive SPM says
there are no significant amplitudes anywhere in $x$. Note however
that this is only an heuristic argument since a peaked
configuration space packet runs counter to the above resonance
approximation (peaked momentum distribution). There are also other
incongruities in this naive application of the SPM. If one recalls
the well known step case with $E>V_{\0}$, single peak reflection
occurs instantaneously (zero delay time). One might expect that
our results tend to this case in the limit $l\to \infty$. This is
not the case. It is also possible in some off-resonance cases to
find negative "delay times". In these latter cases the maximum of
the reflected wave and incoming wave would exist
contemporaneously. This situation also implies problems with
probability conservation.

Numerical calculations automatically conserve probabilities, at
least to within the numerical errors. So to understand what is
happening we performed such calculations and an example of these
is shown in Fig.\,1, where a {\em complex gaussian} modulation
function
\[ g(k)= \left(\frac{a^{\2}}{8 \, \pi^{\3}}\right)^{\frac{1}{4}}
\exp\left[-\frac{a^{\2} (k -k_{\0})^{\2}}{4}\right]\,
\exp[-ik\,x_{\0}]
\]
has been used. It is to be noted that the choice of including a
phase factor in $g(k)$ simply shifts all times by a constant
$m\,x_{\0}/k_{\0}$ at resonance. These figures display the wave
function in the {\em proximity} of the barrier for suitably chosen
times. One clearly sees in these figures the appearance of
multiple peaks due to the two reflection points at $x=0$ and
$x=l$. This observation suggested the following analysis and
imposed the subsequent interpretation.

The $R$, $A$, $B$ and $T$ amplitudes may be rewritten as series
expansions by considering multiple reflections and transmission in
the potential discontinuity points,
%\[
%\begin{array}{l}
%\begin{array}{r|l c r|l }
%\exp[ikx] +  R_{\1}  \exp[-ikx] & \, A_{\1}  \exp[iqx]&
% & A_{\1} \exp[iqx] \,+\,  B_{\1}  \exp[-iqx]  & \, T_{\1} \exp[ikx]\\
% R_{\2}  \exp[-ikx] & \, A_{\2}  \exp[iqx] \,+\,  B_{\1}  \exp[-iqx]&
% & \hspace*{0.5cm} A_{\2} \exp[iqx] \,+ \, B_{\2}  \exp[-iqx]
%& \, T_{\2} \exp[ikx]\\
% \vdots \hspace*{1.2cm} &\hspace*{.75cm} \vdots \hspace*{.95cm}+ \hspace*{.7cm}
%\vdots \hspace*{1.2cm}&  & \vdots \hspace*{.9cm}+ \hspace*{.7cm}
%\vdots \hspace*{1.2cm}& \hspace*{.6cm}\vdots
%\\
%R_{n}  \exp[-ikx] & \, A_{n} \exp[iqx] +  B_{n - \1} \exp[-iqx]&
% & \hspace*{.5cm} A_{n} \exp[iqx] +  B_{n}  \exp[-iqx]
%& \, T_{n} \exp[ikx]
%\end{array}\\
%\hspace*{3.35cm} x=0 \hspace*{8.75cm} x=l
%\end{array}
%\]
%The  continuity constraints of $\Psi(x,t)$ and its derivative at
%$x=0$ and $x=l$ determines the coefficients $R_{n}$, $A_{n}$,
%$B_{n}$, and $T_{n}$,
\begin{eqnarray}
R      &=&  \sum_{n = \1}^{\infty}R_{n} =
 R_{\1} + R_{\2}  \left[ 1 - \left(
\frac{k - q}{k + q}\right)^{\2} \, \exp [2\, iql\,] \right]^{- \1}\, \, ,\nonumber \\
 A
&=& \sum_{n = \1}^{\infty}A_{n} = A_{\1}\left[ 1 - \left( \frac{k
- q}{k + q}\right)^{\2} \, \exp [2\, iql\,] \right]^{- \1}\, \, ,\nonumber\\
 B
&=& \sum_{n = \1}^{\infty}B_{n} = B_{\1}\left[ 1 - \left( \frac{k
- q}{k + q}\right)^{\2} \, \exp [2\, iql\,] \right]^{- \1} \, \, ,\nonumber\\
 T
&=& \sum_{n = \1}^{\infty}T_{n}\, = \, T_{\1} \left[ 1 - \left(
\frac{k - q}{k + q}\right)^{\2} \, \exp [2\, iql\,] \right]^{-
\1}\, \, ,
\end{eqnarray}
with
\begin{eqnarray}
R_{\1} = \frac{k - q}{k + q}\, \, , \, \, \, \,  A_{\1}=
\frac{2\,k}{k + q}\, \, , \, \, \, \, B_{\1}= \frac{2\,k (q -
k)}{(k + q)^{\2}}\, \exp[2 \,iql\,]\, \, , \, \, \, \, T_{\1}=
\frac{4 \,k  q}{(k + q)^{\2}}\,\exp[i\,(q - k)l\,]\, \, ,\nonumber\\
R_{\2} =\frac{q}{k}\, A_{\1} \, B_{\1}\, \, ,\hspace*{11,3cm}\\
\frac{R_{n + \2}}{R_{n + \1}} = \frac{A_{n+ \1}}{A_{n}} =
\frac{B_{n + \1}}{B_{n}} = \frac{T_{n + \1}}{T_{n}} =\left(
\frac{k - q}{k + q}\right)^{\2} \, \exp [2\, iql\,] \hspace*{1cm}
\mbox{\small $n=1,2,\dots$}\, \, .\hspace*{2.4cm}\nonumber
\end{eqnarray}
These sums reproduce exactly the expressions in Eq.(\ref{cexp}).
In this form the interpretation is easy. $R_{\1}$ represents the
first reflected wave (it has no time delay since it is real).
$R_{\2}$ represents the second reflected wave. As a consequence of
continuity, it is the sum, in region II,  of the first left-going
wave ($B_{\1}$) and the second right-going amplitude $(A_{\2})$,
i.e.,
\[ R_{\2}= A_{\2} + B_{\1} \equiv \frac{q}{k}\, A_{\1} \,
B_{\1}\, \, .\] This structure is that given by considering two
``step functions'' back-to-back. Thus at each interface the
``reflected'' and ``transmitted'' waves are instantaneous i.e.
without any delay time. Indeed the SPM {\em applied separately} to
each term in the above series expansion for $R$ yields delay times
which are integer multiples of $2\,(\mbox{d}q/\mbox{d}E)_{\0}l=
2\, (m /q_{\0}) l$. This agrees perfectly with the fact that since
the peak momentum in region II is $q_{\0}$, the $A$ and $B$ waves
have group velocities of $q_{\0}/m$ and hence transit times (one
way) of $(m/q_{\0})l$. The first transmitted peak appears
(according to this version of the SPM) after a time $(m/q_{\0})l$,
in perfect accord with the above interpretation.

Let us re-express what is happening. The incoming wave peak
reaches the first potential discontinuity at $x = 0$. It
instantaneously yields a first reflected peak ($R_{\1}$) and
right-moving ($A_{\1}$) peak in region II. When this later wave
packet reaches at time $t= (m/q_{\0})l$ the second discontinuity
at $x = l$, a part $T_{\1}$ is transmitted into region III ($x >
l$) while a part $B_{\1}$ is turned back and eventually gives rise
to the second reflected peak and so forth. Is this compatible with
probability conservation? It is because of the following identity
\begin{equation}
\sum_{n = \1}^{\infty} \left( |R_{n}|^{\2} + |T_{n}|^{\2} \right)
=1\, \, . \end{equation}
This result is by no means obvious since
it coexists with the well known result, from the plane wave
analysis,
\begin{equation}
|R|^{\2} + |T|^{\2} = |\sum_{n = \1}^{\infty}R_{n} \,|^{\2} + |
\sum_{n = \1}^{\infty}T_{n} \,|^{\2}= 1\, \, .
\end{equation}
In Fig.\,2 we have re-plotted for various times the numerical
calculations displayed in Fig.\,1  and also the separate integral
calculations based upon the above multiple pole model i.e. for
particular $R_{n}$($T_{n}$). The latter wave packets are
represented by the curves. The former un-decomposed numerical
calculations are plotted by various bullets. Agreement is
excellent.

In conclusion, the results of the SPM depend critically upon the
manipulation of the amplitude prior to the application of the
method. A posteriori this seems obvious. If we consider an
amplitude say
\[z(k;x,t)=|z|\exp[i\alpha]\]
 the SPM will yield one peak position for each
given time. If we write the identity
\[z = z_{\1} + z_{\2} \, \, , \]
where $z_{\1}=z - w$ and  $z_{\2}=w$,
 and treat separately these terms, then the same approach
 will yield two peaks and so forth. The method is inherently ambiguous
 unless we know, by some other means, at least the number of
separate peaks involved. Our above barrier analysis is simply a
particular example of this ambiguity, for which we have presented
a simple resolution, based upon multiple reflections, confirmed in
detail by numerical calculations.

\newpage

\begin{figure}[hbp]
\begin{center}
\hspace*{-2.5cm}
\includegraphics[width=20cm, height=22cm, angle=0]{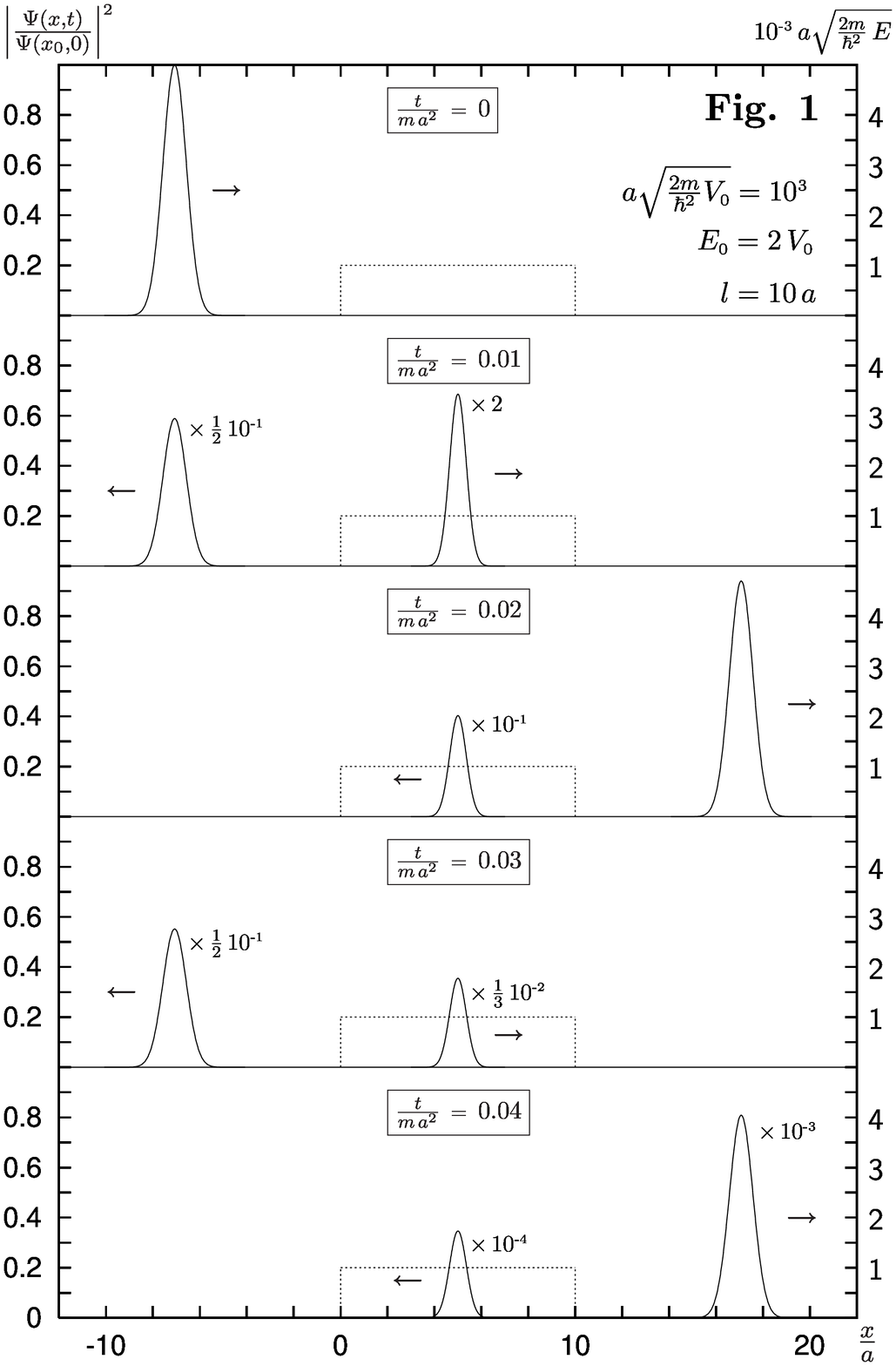}
\caption{The square of the amplitude modulus for different time
frames. Only a fixed region in $x$ close to the barrier is shown.
Each figure should be multiplied by the adjacent factor, where it
exists, to obtain the true curve. The parameters chosen for the
plot are listed in the first frame.}
\end{center}
\end{figure}

\newpage

\begin{figure}[hbp]
\begin{center}
\hspace*{-1.5cm}
\includegraphics[width=14cm, height=18cm, angle=90]{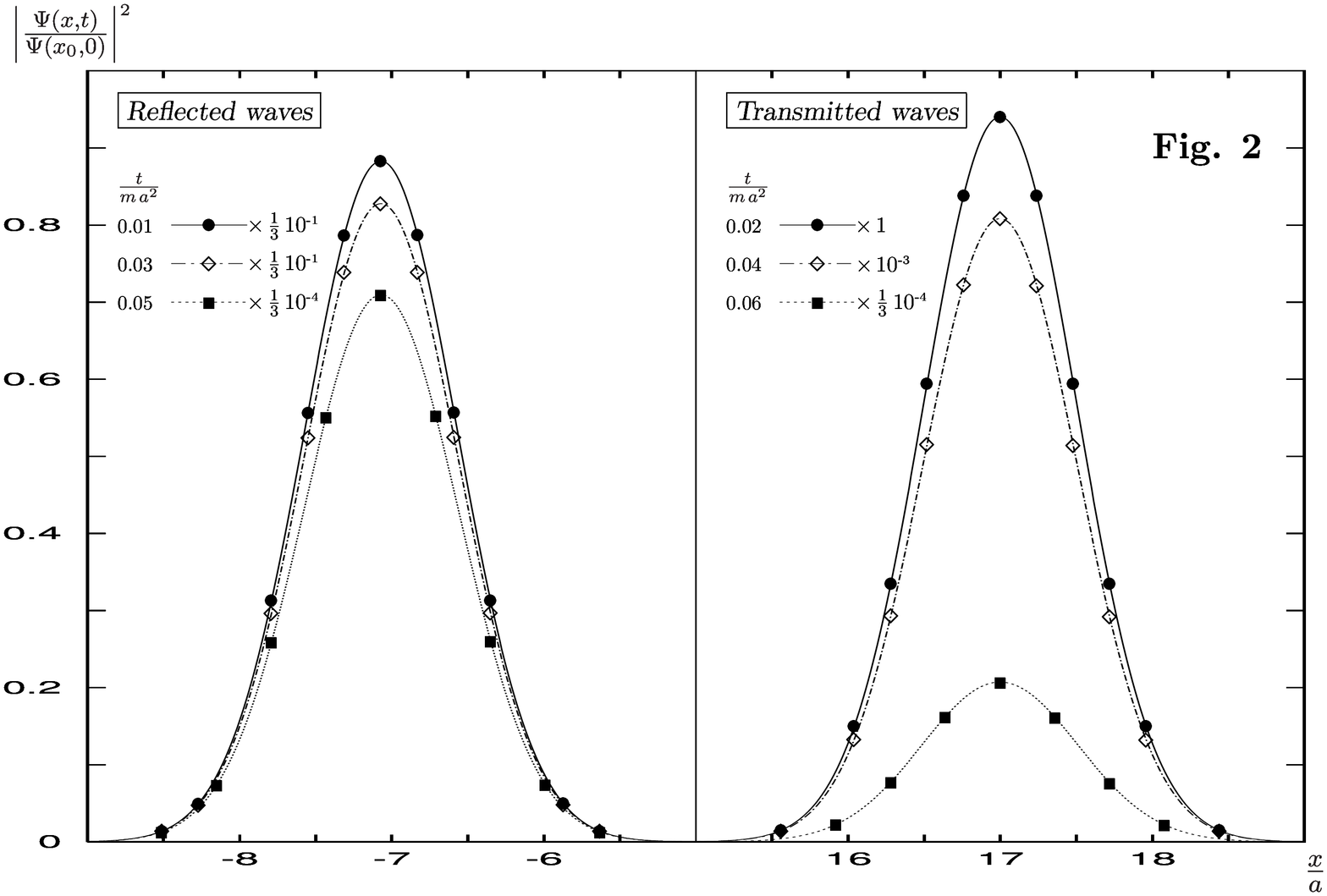}
\caption{Plots of the first few reflected and transmitted waves at
corresponding times. The bullets are from the numerical
convolution of the plane-wave solution. The curves are from the
separate convolution integrals of the first three  $R_{n}$ and
$T_{n}$. Again the true figures are obtained by multiplying by the
listed factors.}
\end{center}
\end{figure}

\end{document}